\documentclass[conference]{IEEEtran}
\IEEEoverridecommandlockouts

\usepackage{cite}
\usepackage{amsmath,amssymb,amsfonts}
\usepackage{algorithmic}
\usepackage{graphicx}
\usepackage{textcomp}
\usepackage[center]{subfigure}
\usepackage{multirow}
\usepackage[
             pdfstartview=FitH,
             plainpages=false,
             bookmarksopen=true,
             bookmarksnumbered=true,
             colorlinks=true,
             bookmarkstype=toc]{hyperref} 

\usepackage{xcolor} 

\graphicspath{{img/}}

\def\BibTeX{{\rm B\kern-.05em{\sc i\kern-.025em b}\kern-.08em
    T\kern-.1667em\lower.7ex\hbox{E}\kern-.125emX}}

\title{Drifting perceptual patterns suggest prediction errors fusion rather than hypothesis selection: \\replicating the rubber-hand illusion on a robot
\thanks{This work has been supported by SELFCEPTION project (www.selfception.eu) European Union Horizon 2020 Programme (MSCA-IF-2016) under grant agreement n. 741941 and the ENB Master Program in Neuro-Cognitive Psychology at Ludwig-Maximilians Universit\"at. Video to this paper: http://web.ics.ei.tum.de/\texttildelow pablo/rubberICDL2018PL.mp4}
\thanks{Accepted for publication at 2018 IEEE International Conference on Development and Learning and Epigenetic Robotics}
}

\author{\IEEEauthorblockN{Nina-Alisa Hinz\IEEEauthorrefmark{2}, Pablo Lanillos\IEEEauthorrefmark{3}\IEEEauthorrefmark{1}, Hermann Mueller\IEEEauthorrefmark{2}, Gordon Cheng\IEEEauthorrefmark{3}}
\IEEEauthorblockA{\IEEEauthorrefmark{2} \textit{Experimental Neuro-Cognitive Psychology}, Ludwig-Maximilians-Universit\"at,
M\"unchen, Germany \\
\IEEEauthorrefmark{3} \textit{Institute for Cognitive Systems}, Technische Universit\"at M\"unchen, M\"unchen, Germany\\
\IEEEauthorrefmark{1}Email: p.lanillos@tum.de
}
}
      
\begin{document}

\maketitle

\begin{abstract}

Humans can experience fake body parts as theirs just by simple visuo-tactile synchronous stimulation. This body-illusion is accompanied by a drift in the perception of the real limb towards the fake limb, suggesting an update of body estimation resulting from stimulation.
This work compares body limb drifting patterns of human participants, in a rubber hand illusion experiment, with the end-effector estimation displacement of a multisensory robotic arm enabled with predictive processing perception. Results show similar drifting patterns in both human and robot experiments, and they also suggest that the perceptual drift is due to prediction error fusion, rather than hypothesis selection.
We present body inference through prediction error minimization as one single process that unites predictive coding and causal inference and that it is responsible for the effects in perception when we are subjected to intermodal sensory perturbations.

\end{abstract}



\section{Introduction}
\label{sec:intro}

Distinguishing between our own body and that of others is fundamental for our understanding of the self. By learning the relationship between sensory and motor information and integrating them into a common percept, we gradually develop predictors about our body and its interaction with the world \cite{lanillos2017enactive}. This body learning is assumed to be one of the major processes underlying embodiment. Body-ownership illusions, like the rubber hand illusion \cite{botvinick1998rubber}, are the most widely used methodology to reveal information about the underlying mechanisms, helping us to understand how the sensorimotor self is computed. Empirical evidence has shown that embodiment is flexible, adaptable, driven by bottom-up and top-down modulations and sensitive to illusions.

We replicated the passive rubber hand illusion on a multisensory robot and compared it with human participants, therefore gaining insight into the perceptive contribution to self-computation. Enabling a robot with human-like self-perception \cite{lanillos2016yielding} is important for: i) improving the machine adaptability and providing safe human-robot interaction, and ii) testing computational models for embodiment and obtaining some clues about the real mechanisms.
Although some computational models have already been proposed for body-ownership, agency and body-illusions, the majority of them are restricted to the conceptual level or simplistic simulation \cite{kilteni2015over}. Examining real robot data and using body illusions as a benchmark for testing the underlying mechanism enriches the evaluation considerably. To the best of our knowledge, this is the first study replicating the rubber hand illusion on an artificial agent and comparing it to human data.

We showed that when inferring the robot body location through prediction error minimization \cite{friston2005theory}, the robot limb drifting patterns are similar to those observed in human participants. Human and robot data suggest that the perception of the real hand and the rubber hand location drifts to a common location between both hands. This supports the idea that, instead of selecting one of two hypotheses (common cause for stimulation vs. different causes) \cite{Samad.2015}, visual and proprioceptive information sources are merged generating an effect similar to averaging both hypotheses \cite{Erro.2018}.

The remainder of this paper is as follows: Sec. \ref{sec:method} describes current rubber-hand illusion findings and its neural basis; Sec. \ref{sec:model} defines the computational model designed for the robot; Sec. \ref{sec:setup} describes the experimental setup for both human participants and the robot; Sec. \ref{sec:results} presents the comparative analysis of the drifting patterns; Finally, Sec. \ref{sec:discussion} discusses body estimation within the prediction error paradigm as the potential cause of the perceptual displacement.

\section{Background}
\label{sec:method}

\subsection{Rubber-hand illusion}

\begin{figure}[htbp!]
	\subfigure[Illusion depending on distance]{\includegraphics[width=0.46\columnwidth, height=100px]{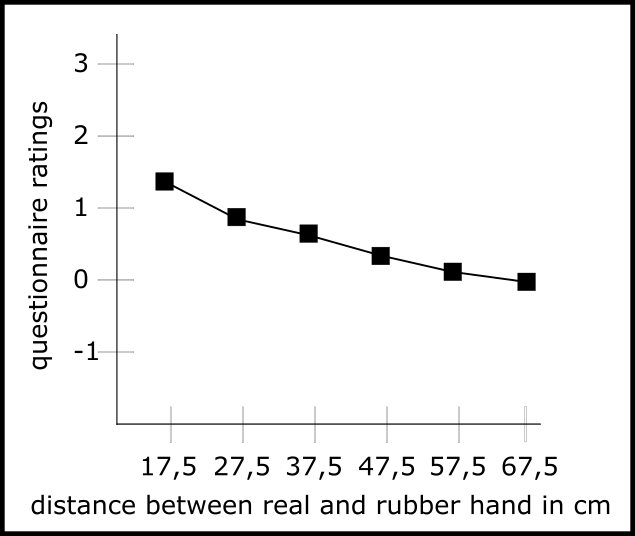} \label{fig:overview:illusionDistance}}
	\subfigure[Proprio. drift depending on time]{\includegraphics[width=0.46\columnwidth, height=100px]{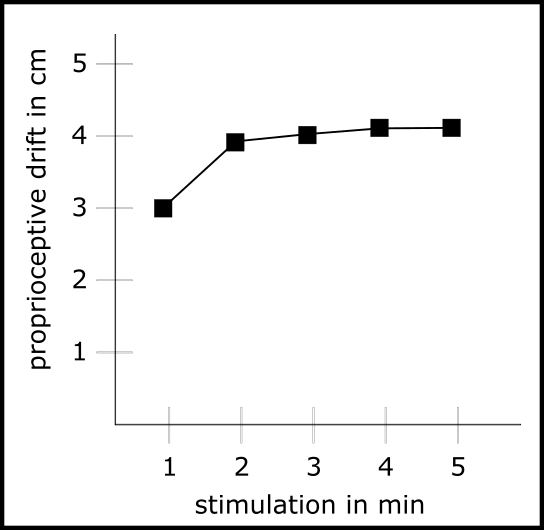} \label{fig:overview:proprioceptiveDrift}}
	\caption{(a) Quality of the illusion as a function of the distance between hands: the higher the better. (b) Intensity of the proprioceptive drift depending on the visuo-tactile stimulation time. Adapted from \cite{Lloyd.2007} and \cite{Tsakiris.2005} respectively.}
	\label{fig:overview}
\end{figure}

Botvinick and Cohen \cite{Botvinick.1998} demonstrated that humans can embody a rubber hand only by means of synchronous visuo-tactile stimulation of the rubber hand and their hidden real hand. This was measured using a questionnaire about the illusion, but also by proprioceptive localization of the participant's real hand. After experiencing the illusion, the perception of their own hand's position had drifted towards that of the rubber hand. Since then, multiple studies have replicated the illusion under different conditions (for a review, see \cite{Kilteni.2015}). Collectively, these studies show that top-down expectations about the physical appearance of a human hand, resulting from abstract internal body models, and bottom-up sensory information, especially spatiotemporal congruence of the stimulation and distance between the hands, influence embodiment of the fake hand \cite{Tsakiris.2005}. In \cite{AymerichFranch.2017}, they were even able to induce body-ownership on a robotic arm. The common assumption is that the spatial representations of both hands are merged, as a result of minimizing the error between predicted sensory outcomes from seeing the stimulation of the rubber hand and the actual sensory outcome of feeling the stimulation of one's own hand \cite{Tsakiris.2010}. Recently, \cite{Erro.2018} undermined this by showing that not only the perception of the real hand's position is drifted towards the rubber hand (proprioceptive drift), but the one of the rubber hand is drifted towards the real hand as well (visual drift), i.e. to a common location between both hands. In \cite{Samad.2015}, they proposed a Bayesian Causal Inference Model for this prediction error minimization, considering visual, tactile and proprioceptive information, weighted according to their precision. In combination with the prior probability of assuming a common cause or different causes for the conflicting multi-sensory information, the posterior probability of each hypothesis is computed. A common cause, i.e. ownership of the rubber hand, is assumed if the posterior probability exceeds a certain threshold. This binarity of the illusion, however, is at variance with findings of \cite{Tsakiris.2005}, demonstrating a continuous proprioceptive drift of the stimulated hand. The proprioceptive drift was shown to increase exponentially during the first minute of stimulation and increasing further over the following four minutes (Fig. \ref{fig:overview:proprioceptiveDrift}). Although the reported onset of the illusion ranged from 11 seconds \cite{Ehrsson.2004} to 23 seconds, with 90 percent of subjects experiencing it within the first minute of stimulation \cite{Kalckert.2017}, the ongoing drifting suggests a continuous, rather than a discrete mechanism, being involved in embodiment.

The proposed computational models in the literature of body-ownership illusions need further verification from experimental data. Several studies showed reduced illusion scores for larger, as compared to smaller, distances between the real and the rubber hand \cite{Lloyd.2007,Zopf.2010,Preston.2013,Pritchard.2016,Ratcliffe.2017} (Fig. \ref{fig:overview:illusionDistance}), though only few studies measured the proprioceptive drift in dependence on the distance between the two hands \cite{Zopf.2010}, \cite{Preston.2013}. While \cite{Zopf.2010} found an increased proprioceptive drift for a larger distance, the relative amount of drift (i.e. corrected for distance) did not differ between the small and large distances. In \cite{Preston.2013}, they replicated this result, as long as the fake hand was near the body. If the real hand was closer to body midline than the fake hand, increasing distance between both hands decreased the proprioceptive drift. 

In the present study, we systematically varied the distance between both hands. The real hand, however stayed in the same position for all conditions and only the fake hand had a varying distance from the real hand in anatomically plausible positions. In \cite{Lloyd.2007}, where the distance between both hands was varied by displacing the fake hand in relation to the real hand, the fake hand was also increasingly rotated with increasing distance. Rotational differences, nevertheless, may influence the illusion \cite{Makin.2008}, probably confounding the results of \cite{Lloyd.2007}. In the current study, we systematically examined the influence of the distance between the rubber and the real hand on proprioceptive and visual drift. This provided the basis for validating the computational model proposed in Sec. \ref{sec:model} and comparing the drift of the body estimation in different distances between the robot and humans.


\subsection{Body illusions in the brain}

A seminal contribution to possible neuronal mechanisms underlying the rubber hand illusion came from \cite{graziano2000coding}, who discovered parietal neurons in the primate brain coding for the position of the real arm and a plosturally plausible fake monkey arm. Several fMRI studies looked into the neural correlates for body-ownership illusions in humans (see \cite{Makin.2008} for a review). Three areas were consistently found activated during the rubber hand illusion: posterior parietal cortex (including intra-parietal cortex and temporo-parietal junction), premotor cortex and lateral cerebellum. The cerebellum is assumed to compute the temporal relationship between visual and tactile signals, thus playing a role in the integration of visual, tactile and proprioceptive body-related signals \cite{Ehrsson.2005}, \cite{Guterstam.2013}. The premotor and intra-parietal cortex are multisensory areas, also integrating visual, tactile and proprioceptive signals present during the rubber hand illusion \cite{Guterstam.2016}. In \cite{Makin.2008}, they differentiated the role of posterior parietal cortex, being responsible for the recalibration of visual and tactile coordinate systems into a common reference frame, and the role of premotor cortex, being responsible for the integration of signals in this common, hand-centered reference frame. Although it is known that these areas participate in evoking the rubber hand illusion, little is known about the underlying computations \cite{Apps.2014}. In \cite{Zeller.2016}, they used dynamic causal modeling during the rubber hand illusion to confirm to some extent that, during the illusion, visual information is weighted more than proprioceptive information - which would be congruent with predictive coding models. During the illusion the intrinsic connectivity of the somatosensory cortex was reduced, indicative of less somatosensory precision, while the extrinsic connectivity between the lateral occipital complex and premotor cortex was increased, indicative of more attention to visual input. Further functional evidence for the proposed computations is needed.

\section{Computational model}
\label{sec:model}

\begin{figure}[htbp!]	
\includegraphics[width=0.9\columnwidth]{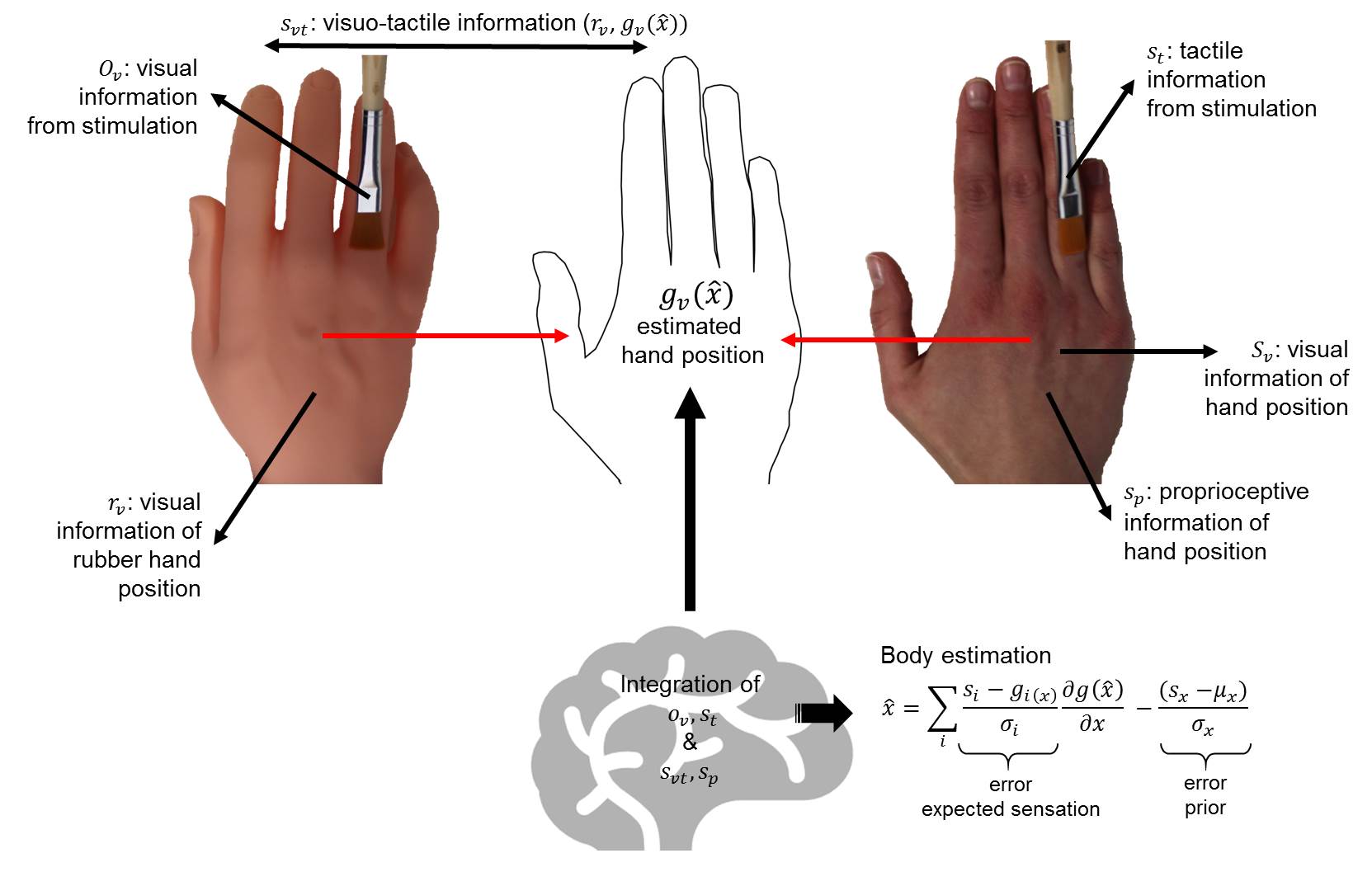} 
\caption{Rubber hand illusion modelled as a body estimation problem solved using prediction error minimization. Visual features of the rubber hand are incorporated when there is synchronous visuo-tactile stimulation, though it is constrained by the prior belief and the expected location of the hand according to the generative visual forward model and the estimated body configuration.}
\label{fig:model}
\end{figure}

We formalized the rubber hand illusion as a body estimation problem under the predictive processing framework. The core idea behind this is that all features and sensory modalities are contributing to refine body estimation through the minimization of the errors between sensations and predictions \cite{lanillos2018adaptive}. During synchronous visuo-tactile stimulation, the most plausible body configuration is perturbed due to the merging of visual and proprioceptive information. This is coherent with the drift of both the real hand and the rubber hand as the participants are just pointing to the prediction of their hand according to the current body configuration. Figure \ref{fig:model} shows how sensory modalities or features are contributing to the estimation of the participant´s limb.

We define $x$ as the latent space variable that expresses the body configuration. We model the problem as inferring the most plausible body configuration $\hat{x}$ given the sensation likelihood and the prior: $P(\hat{x}|s) = p(s|\hat{x}) p(\hat{x})$. We further define $s_p$, $s_v$, $s_{vt}$ as the proprioceptive, visual and visuo-tactile sensation respectively. Assuming independence of the different sources of information we get:

\begin{align}
P(\hat{x}|s) = p(s_p|x)p(s_v|x)p(s_{vt}|x) p(x)
\end{align}

The perception or estimation of the body is then solved by learning an approximation of the forward model for each feature or modality $s =g(x)$ and minimizing a lower bound on the KL-divergence known as negative free energy F \cite{friston2005theory,bogacz2015tutorial}. 
\begin{align}
 \log P(\hat{x}|s) = -F =  \sum_i \log p(s_i|x) + \log p(x)
\end{align} 
We obtain the estimated value of the latent variables through gradient descent minimization  $\hat{x}= \frac{\partial F}{x}$:

\begin{align}
	\hat{x} = \sum_i \underbrace{\frac{(s_i - g_i(\hat{x}))}{\sigma_{s_i}}}_{\text{error expected sensation}} \!\!\!\!\!\!g'_i(\hat{x}) - \underbrace{\frac{s_x- \mu_x}{\sigma_x}}_{\text{error prior}}
\end{align}
We assume that all sensations / features follow a Gaussian distribution with (linear or non-linear) mean $g_i(x)$ and variance $\sigma_{s_i}$. The forward models learned should be differentiable with respect to the body configuration ($g'_i(\hat{x}) = \partial g_i(\hat{x})/\partial x$).  

By rewriting the prediction error as $e = s-g(x)$ and defining $\mu_x$ as the prior belief about the body configuration, the dynamics of the body perception model are described by (see Appendix for derivation and \cite{lanillos2018adaptive} for the detailed algorithm):
\begin{align}
&\dot{x} = -e_x + e_p  + e_v g'_v(\hat{x}) + e_{vt} g'_{vt}(\hat{x})\nonumber\\
&\dot{e}_x = s_x  -  \mu_x  - \sigma_x e_x \nonumber\\
&\dot{e}_p = s_p  -  \hat{x} - \sigma_p e_p\nonumber\\
&\dot{e}_v = s_v  -  g_v(\hat{x}) - \sigma_v e_v\nonumber\\
&\dot{e}_{vt} = s_{vt}  -  g_{vt}(\hat{x}) - \sigma_{vt} e_{vt}\nonumber\\
&\dot{\mu}_x = \mu_x + \lambda e_x
\end{align}
where $\lambda$ is the learning ratio parameter that specifies how fast the prior of the body configuration $\mu_x$ is adjusted to the prediction error. 

The visual forward function $g_v$ and its derivative are calculated using Gaussian process estimation (see Sec. \ref{sec:setup}). The visuo-tactile generative function is computed by means of a hand-crafted likelihood, which uses the visual $o_v$ and tactile $s_t$ stimulation information (temporal $h_s$ and spatial $h_t$), and the expected position of the hand $g_v(\hat{x})$:
\begin{align}
\label{eq:touch}
g_t(\hat{x}) = s_t \cdot h_s \cdot h_t = s_t  a_1 e^{-b_1\sum(g_v(\hat{x}) - o_v)^2} \cdot a_2e^{-b_2 \delta^2}
\end{align}
where $a_1,b_1,a_2,b_2$ are parameters that shape the likelihood of the spatial plausibility and have been tuned in accordance with the data acquired from human participants; $\delta$ is the level of synchrony of the events (e.g. timing difference between the visual and the tactile event); and $o_v$ is the other agent end-effector location in the visual field.

\section{Experimental Set-up}
\label{sec:setup}

\subsection{Participants selection}
20 volunteers (mean age: 25, 75 \% female) took part in the experiment. They received 8 euros per hour in compensation. All participants were right-handed, had no disability of perceiving touch on their right hand, did not wear nail polish and did not have any special visual features (e.g. scars / tattoos) on their right hand. They had no neurological or psychological disorders, as indicated by self-report, and normal or corrected-to-normal vision. None of them had experienced the rubber hand illusion before. All participants gave informed consent prior to the experiment.


\subsection{Humans experiment details}

\begin{figure*}[htbp!]
\centering
\subfigure[Human]{\includegraphics[width=0.44\textwidth, height=110px]{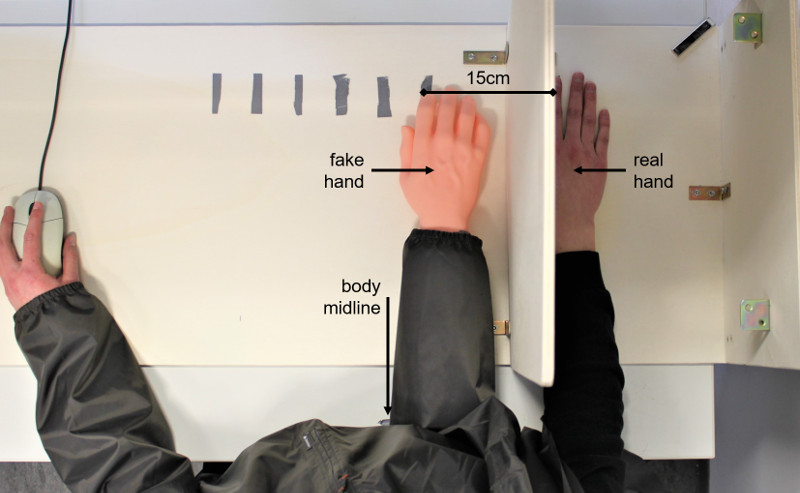} \label{fig:setup:1}}
\subfigure[Robot]{\includegraphics[width=0.44\textwidth, height=110px]{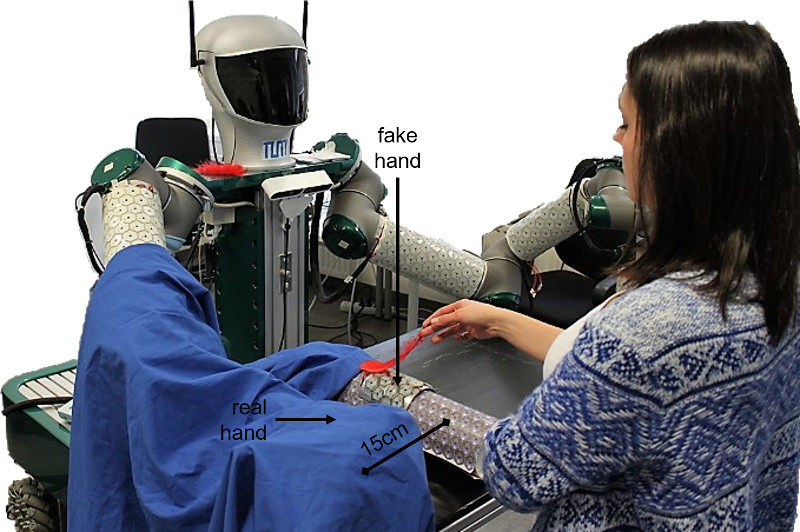} \label{fig:setup:2}}
\caption{Experimental setups used in the current study. (a) Rubber-hand illusion in different distance conditions and (b) artificial version.}
\label{fig:setup}
\end{figure*}

We performed the rubber-hand illusion experiment, focusing on the proprioceptive drift of the real and the visual drift of the rubber hand, as a function of the distance between both hands. The experiment, depicted in Fig. \ref{fig:setup}, comprised six conditions, each with a different distance between the real hand and the rubber hand. The participant's real right hand was placed in a box, with the index finger 20cm away from the participant’s body midline. The rubber hand was placed with its index finger 15cm, 20cm, 25cm, 30cm, 35cm or 40cm away from the participant’s real right hand (5cm, 0cm, -5cm, -10cm, -15cm or -20cm away from the participant’s body midline respectively).

Participants sat in front of a wooden box, placing their hands near the outer sides of the box. They wore a rain cape covering their body and arms. In one of the arms, a rubber hand was placed such that it seemed coherent with the body. With their left hand they held a computer mouse. Each trial consisted of four phases: localization of the real hand, localization of the rubber hand, the stimulation phase and post-stimulation localization of both hands. 
\begin{enumerate}
\item First, we covered the box with a wooden top and a blanket above it, so that no visual cues could be used. Participants had to indicate where they currently perceived the location of the index finger of their right hand, pointing with the mouse on a horizontal line presented on the screen. The line did cover the whole length of the box.
\item After fixating the rubber hand for one minute, we again covered the box and the same task was repeated for the rubber hand.  
\item The box was remodeled, removing the cover and introducing a vertical board next to the participant's right hand so that it was not visible to the participant (Fig. \ref{fig:setup:1}).  Then the experimenter began stimulating the rubber and the real hand synchronously with two similar paintbrushes, starting at the index finger, continuing to the little finger and then starting at the index finger again, with one brush stroke each about two seconds. 
\item Subsequently, participants were again asked to indicate where they perceived the index finger of the real or the rubber hand, starting with the real or the rubber hand in randomized fashion. The box was covered during the localization.
\item At the end of each trial, participants were asked to answer ten questions related to the illusion adapted from \cite{Kammers.2009}, presented randomized on the screen, using a continuous scale from -100 (indicating strong disagreement) to 100 (indicating strong agreement).
\end{enumerate}
For the localization trials, a horizontal line was presented on the screen opposite to the box, with the screen having the same size as the box. The localization trials were repeated ten times to account for high variance. The proprioceptive drift and visual drift were calculated by subtracting the average of the first localization phase from the second localization phase for the real hand and the rubber hand separately. The illusion index was calculated by subtracting the average response to control statements S4-S10 from the average response to illusion statements S1-S3 \cite{Abdulkarim.2016}. Between all phases participants were blindfolded, so they did not observe the remodeling, which might potentially have impeded the illusion.

\subsection{Robot experiment details}

We tested the model on the multisensory UR-5 arm of robot TOMM \cite{dean2017tomm}, as depicted in Fig. \ref{fig:setup:2}. The proprioceptive input data were three joint angles with added noise (shoulder$_1$, shoulder$_2$ and elbow - Fig. \ref{fig:data:a1}). The visual input was an rgb camera mounted on the head of the robot, with $640\times480$ pixels definition. The tactile input was generated by multimodal skin cells distributed over the arm \cite{mittendorfer2011humanoid}.

\subsection{Learning $g(x)$ from visual and proprioceptive data}
\begin{figure}[!hbtp]
\centering
\subfigure[Data recorded example (joints angles + noise, end-effector visual, end-effector cartesian) and schematic picture of the 3-DOF.]{\includegraphics[width=0.90\columnwidth, height=120px]{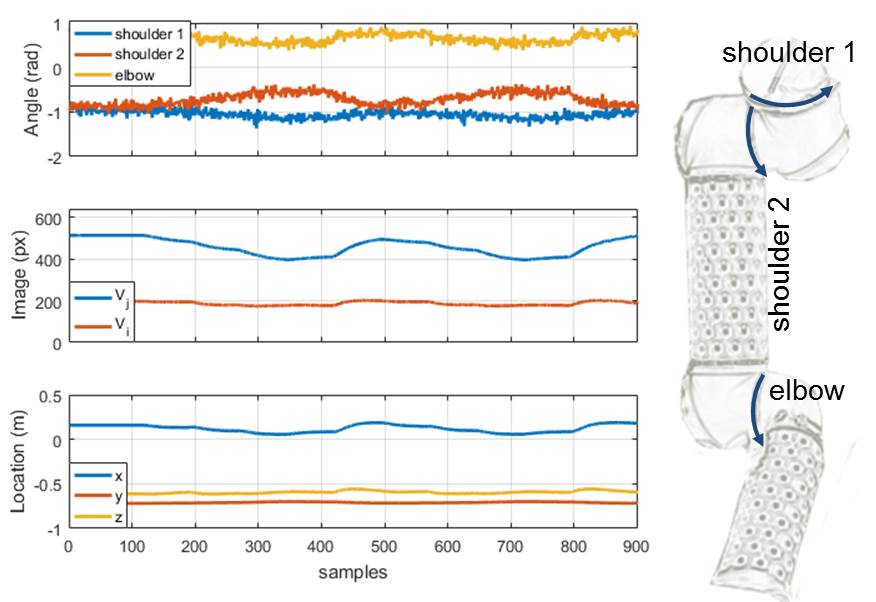} 
\label{fig:data:a1}}\\
\subfigure[Learnt $g_v(x)$ for visual horizontal location]{\includegraphics[width=0.90\columnwidth, height=85px]{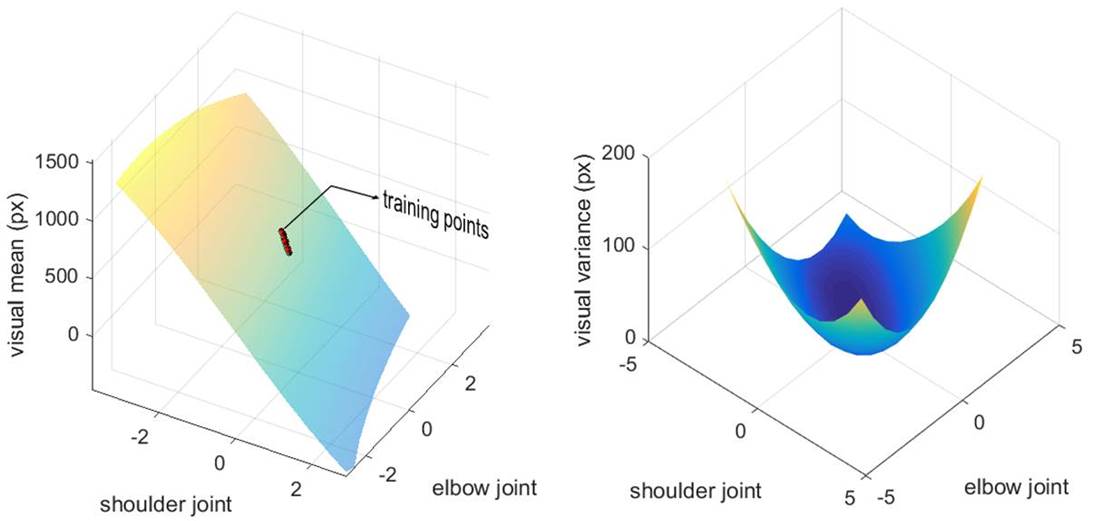}
\label{fig:data:a2}}\\
\subfigure[Tactile (left) and visual (right) event trajectories]{\includegraphics[width=0.80\columnwidth,height=80px]{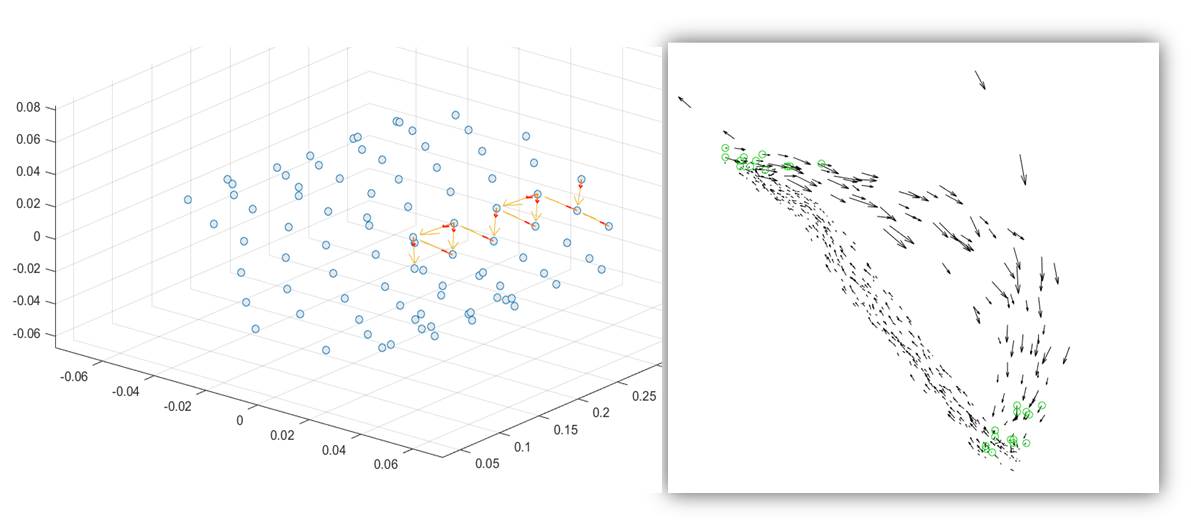} 
\label{fig:sync:a1}}
\caption{Collected data. (a) Joints angles, visual and ground truth information of the end-effector. (b) Mean and the variance computed by the GP for the visual horizontal location depending on two joints. (c) Touch patterns extracted from tactile (117 forearm skin cells) and visual sources.}
\label{fig:data}
\end{figure}

In order to learn the sensory forward model, we applied Gaussian process (GP) regression: $g_v(x) \sim \mathcal{GP}$. We programmed random trajectories in the joint space that resembled horizontal displacements of the arm. Figure \ref{fig:data:a1} shows the extracted data: noisy joint angles and visual location of the end-effector, obtained by color segmentation. To learn the visual forward model $s_v = g_v(x)$, each sample was defined as the input joint angles sensor values $x= (x_1, x_2, x_3)$ and the output $s_v = (i,j)$ pixel coordinates. As an example, Figure \ref{fig:data:a2} shows the learned visual forward model by GP regression with 46 samples (red dots). It describes the horizontal mean and variance (in pixels) with respect to two joints angles. The GP learning and its partial derivative computation with respect to $x$ is described in the Appendix \ref{sec:appendix:gp}.

\subsection{Extracting visuo-tactile data}

We used proximity sensing information (infrared sensors) from 117 different skin cells to discern when the arm was being touched. The sensor outputed a filtered signal $\in (0,1)$. From the other's hand visual trajectory and the skin proximity activation, we computed the level of synchrony between the two patterns (Fig. \ref{fig:sync:a1}). Timings for tactile stimuli $s_t$ were obtained by setting a threshold over the proximity value: prox $> 0.7 \rightarrow$ activation. Timings for the other's trajectory events were obtained through the velocity components. Detected initial and end positions of the visual touching are depicted in Fig. \ref{fig:sync:a1} (right, green circles).

\section{Results}
\label{sec:results}
We compared the drifting data extracted from the rubber-hand illusion experiment in human participants and the robot. In order to obtain the robot results, we fixed in advance the model parameters for the learning and the body estimation stages. $g_v(x)\sim \mathcal{GP}$  learning hyperparameters: signal variance $\sigma_n = \exp(0.02)$ and kernel length scale $l=\exp(0.1)$. The integration step was $\Delta_t= 0.05$ ($20Hz$) and the error variances were $\sigma_x \in \mathcal{R}^3 = [1,1,1]$, $\sigma_p \in \mathcal{R}^3 = [1,1,1]$, $\sigma_{vt} \in \mathcal{Z}^2= [2,2]$. The adaptability rate of $\mu_x$ was $\lambda = 1$. The visual feature from the real hand $s_v$ was not used in the rubber hand illusion experiment as the arm was covered. Finally, the visuo-tactile function (Eq. \ref{eq:touch}) parameters were: $b_1= \frac{\sigma_t}{d_{max}^2}$, where $\sigma_t=0.001$ and $d_{max}=0.0016$; $b_2 = 25$; and $a_1=a_2=1$. The robot drift was computed by subtracting the estimated end-effector position $g_v(\hat{x})$ and the ground truth location, and $\hat{x}$ was dynamically updated minimizing the prediction error using the proposed model.

\subsection{Comparative analysis}
\begin{figure}[htbp!]
\centering
\subfigure[Drift]{\includegraphics[width=0.48\columnwidth]{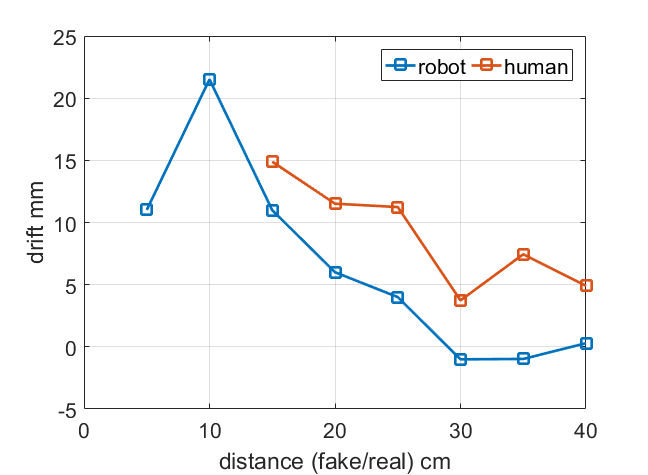} \label{fig:results:1}}
\subfigure[Relative drift]{\includegraphics[width=0.48\columnwidth]{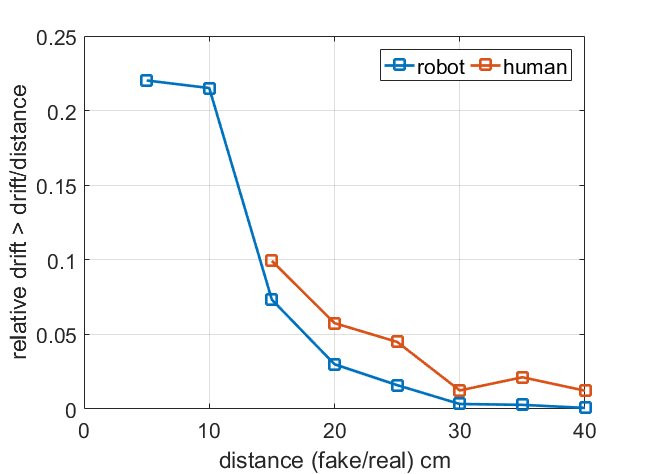} \label{fig:results:2}}
\caption{Human proprioceptive drift vs end-effector robot estimation drift after the rubber hand illusion experiment. (a) Drift in cm. Positive values express displacement towards the fake hand. (b) Relative drift depending on the distance between fake and real arm.}
\label{fig:results}
\end{figure}
Figure \ref{fig:results} shows the proprioceptive drift comparison. Fig. \ref{fig:results:1} shows similar drifting patterns in both the robot and the human participants. A drift towards the fake hand emerges in both cases when the distance is small and then vanishes with longer distances. The prior information used for the tactile likelihood function parameters is modulated when the effect is taking place, as the error will start propagating when the gradient of the function is noticeable. Furthermore, the relative drift (Fig. \ref{fig:results:2}) also showed that, for close distances, the amount of displacement is the same, and then it decreases until vanishing. The robot was tested on even closer distances than humans, since the human experimental setup was not equipped for distances beneath 15cm. The large increase in proprioceptive drift for 10cm distance between fake and real hand is an interesting prediction for human data, that could be tested in future work.


\subsection{Human data analysis}
Data exceeding a range of two standard deviations around the mean was excluded from further analysis. T-tests were used to test the proprioceptive drift, the visual drift and the illusion score in each condition against zero. In the first three conditions the proprioceptive drift was significantly different from zero, while it was not in the other three conditions (Table \ref{tab:ttest}). Employing Bonferroni-Correction only leaves a trend towards significance in the 20cm distance condition. However, the average of the first three conditions is still significantly different from zero $M: 13.72$, $SD: 14.84$, $p <.001$) while the average of the other conditions is not ($M 8.24:$, $SD: 19.79$, $p > .05$). The visual drift was only significantly different from zero in the 30cm distance condition ($M: 14.01$, $SD: 29.72$, $p < .01$).  The illusion score was significantly different from zero in all conditions (all $p < .05$). Partial Pearson correlations between illusion score and proprioceptive drift, illusion score and visual drift and between proprioceptive drift and visual drift were not significant (all $p > .05$).

\begin{table}[htbp]
\caption{Descriptive and inference statistics from proprioceptive drift data in each condition.}
\begin{center}
\begin{tabular}{|l|c|c|c|c|c|}
\hline
\textbf{Condition} & \textbf{mean}& \textbf{std}& \textbf{df} & \textbf{t-value} & \textbf{p-value} \\
\hline
15\,cm & 14.89\,mm & 27.08\,mm & 19 & 2.46 & .024 \\
20\,cm & 12.79\,mm & 18.99\,mm & 17 & 2.86 & .011 \\
25\,cm & 13.22\,mm & 23.51\,mm & 16 & 2.32 & .034\\
30\,cm & 4.39\,mm & 15.52\,mm & 16 & 1.17 & .260\\
35\,cm & 7.84\,mm & 18.00\,mm & 18 & 1.90 & .074\\
40\,cm & 5.45\,mm & 31.34\,mm & 17 & 0.74 & .471\\
\hline
\end{tabular}
\label{tab:ttest}
\end{center}
\end{table}

\begin{figure}[htbp!]
	\centering
	\subfigure[propriceptive drift]{\includegraphics[width=0.48\columnwidth]{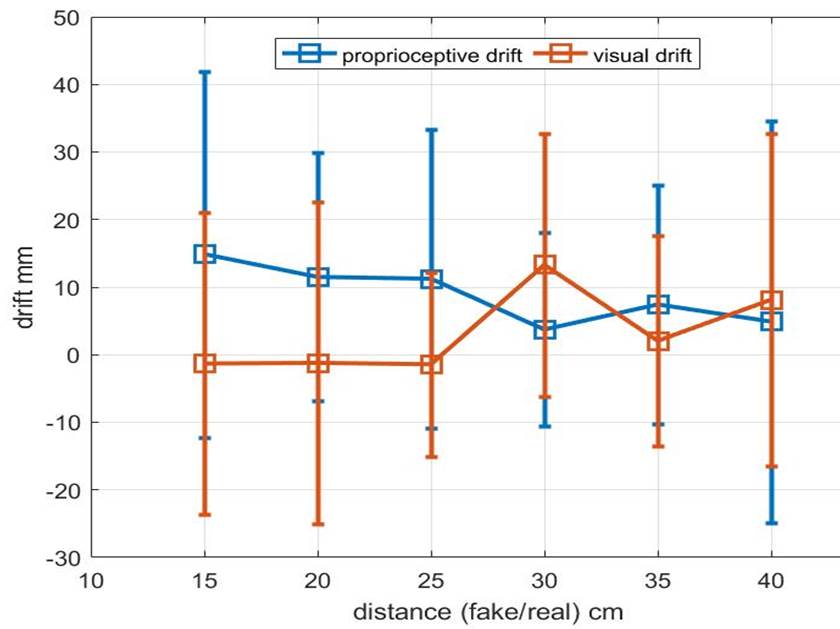} \label{fig:humanResults:1}}
	\subfigure[illusion score]{\includegraphics[width=0.48\columnwidth]{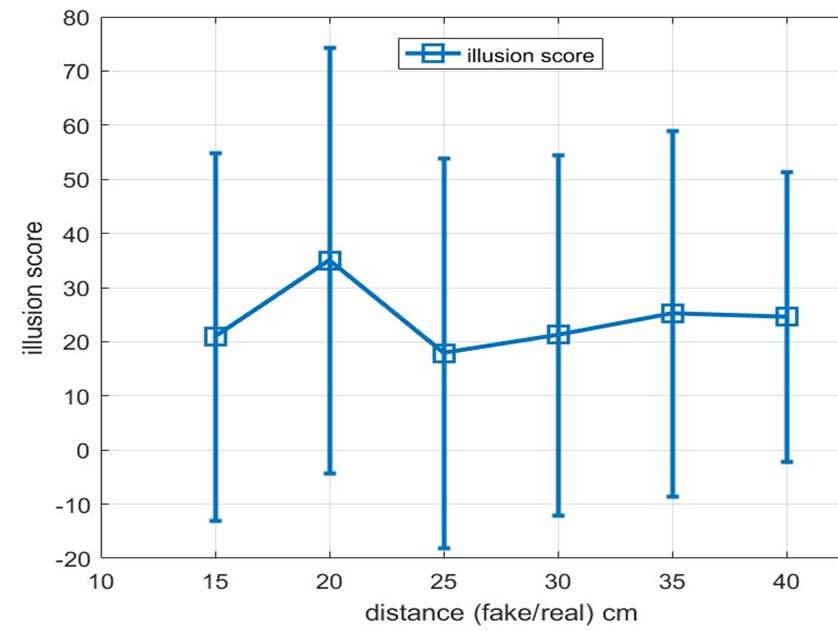} \label{fig:humanResults:3}}
	\caption{Mean and standard deviation of (a) the visual and the proprioceptive drift, and (b) the illusion score depending on the distance.}
	\label{fig:humanResults}
\end{figure}

\subsection{Robot model analysis}

\begin{figure}[!hbtp]
\centering
\subfigure[End-effector drift depending on the sensing information (propio., visuo-tactile, and proprio+visuo-tactile)]{\includegraphics[width=0.95\columnwidth, height=90px]{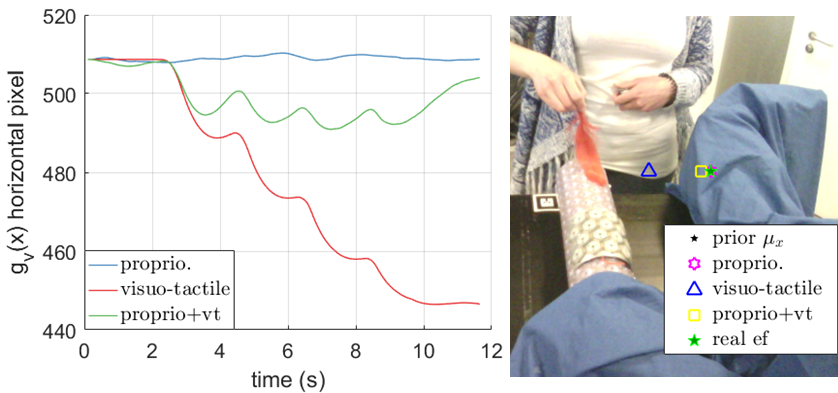} 
\label{fig:resultsrh:a2}}
\subfigure[Evolution of body estimation and predictive errors]{\includegraphics[width=0.99\columnwidth, height=220px]{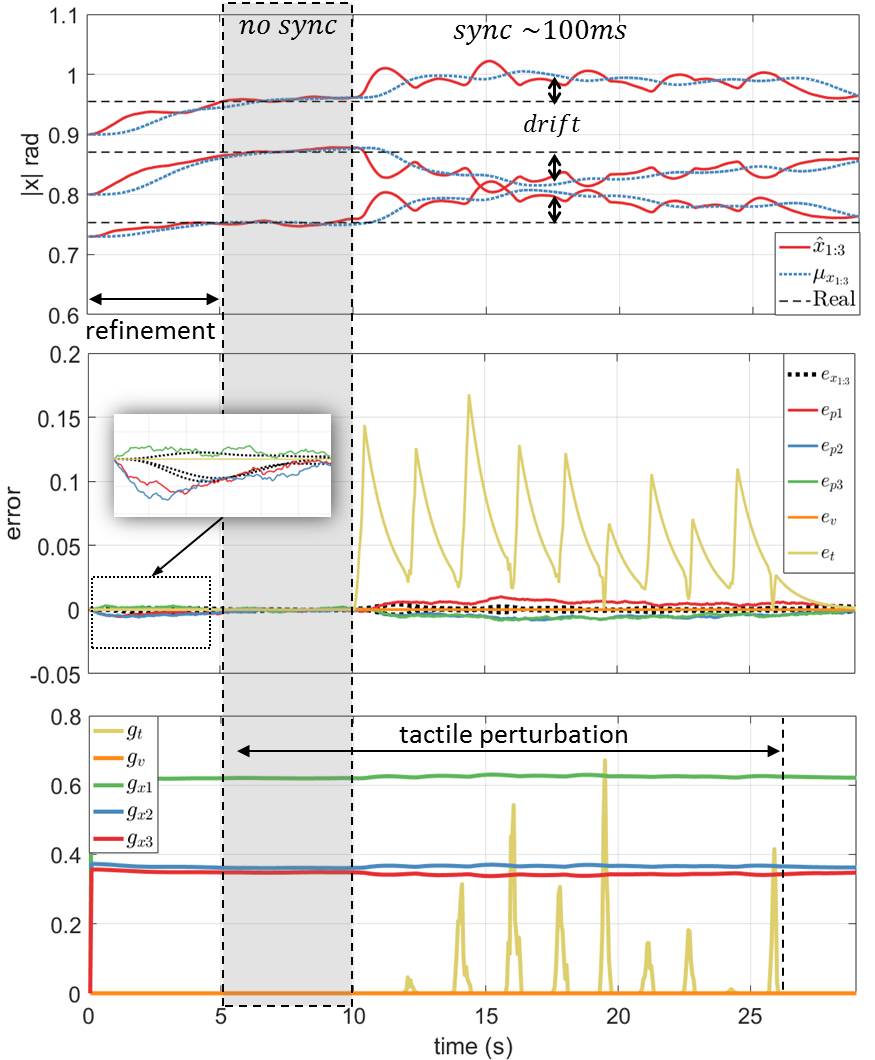}
\label{fig:resultsrh:a3}}
\caption{Replicating the rubber-hand illusion on a robot. (a) End-effector estimation depending on the modality used. (b top) Joints angles in radians: real (black dotted line), estimated $\mu_x$ (blue dotted line) and current belief $\hat{x}$ (red line). (b middle)  Errors between expected and observed values. (b bottom) $g(x)$ values evolution during the experiment. In the first five seconds, in which there is no tactile stimulation, the estimation is refined. Next, we inject tactile stimulation while the experimenter is touching a fake arm. When visuo-tactile stimulation becomes synchronous, a horizontal drift appears on the end-effector estimation.}
\label{fig:resultsrh}
\end{figure}

We analyzed the internal variables of the proposed model during the visuo-tactile stimulation and the induced end-effector estimation drift towards the fake arm. Figure \ref{fig:resultsrh:a2} shows the robot camera view with the final end-effector estimation overlaid after 12 seconds. Depending on the different enabled modalities (proprioceptive, visuo-tactile and proprioceptive+visuo-tactile), body estimation evolved differentially, accordingly the prediction of the end-effector $g_v(x)$. Fig. \ref{fig:resultsrh:a3} shows the evolution of the body configuration in term of joint angles and the corresponding prediction errors. We did initialize the robot belief in a wrong body configuration to further show the adaptability of the model. During the first five seconds, the system converged to the real body configuration. Afterwards, when perturbing with synchronous visuo-tactile stimulation, a bias appeared on the body joints. This implies a drift of the robot end-effector towards the location of the fake arm visual feature. Tactile perturbations are shown as prediction error bumps (yellow line). Fig. \ref{fig:resultsrh:a3}, top plot, also shows how smooth body configuration output $\mu_{x_{1:3}}$ is (blue line). The robot inferred the most plausible body joints angles given the sensory information, which in this case yielded a horizontal drift on the estimated end-effector location. A video of the evolution of the variables during the artificial rubber-hand illusion experiment can be found at \url{http://web.ics.ei.tum.de/~pablo/rubberICDL2018PL.mp4}.

\section{Discussion: Body estimation as an explanation for the perceptual drift}
\label{sec:discussion}

It has been shown that during the rubber hand illusion, the location of the real hand is perceived to be closer to the rubber hand than before. Similarly, the location of the rubber hand is mislocalized towards the real hand \cite{Erro.2018}. Our results from the robot and humans support the former finding: in our predictive coding scheme, the representation of both hands merged into a common location between both hands, due to inferring one's body's location from minimizing free energy. This body estimation generated a drift of the perceived location of the real hand towards the equilibrium location, which was visible in the data from both the humans and the robot. In comparative analysis, the patterns of the drift resembled each other, both in terms of absolute and relative values. The three closest tested distances showed a substantial proprioceptive drift. All other distances showed a smaller drift, approaching zero. For these, the distance between the fake and the real hand was probably too large for the fake hand to be fully embodied, supporting \cite{Samad.2015} simulation data exhibiting a reduced illusion probability for distances over 30cm. Although previous and the present research support predictive coding as a probable underlying mechanism of the rubber hand illusion, other accounts can not be ruled out by the present work.

Human illusion score data, however, did not mirror the proprioceptive shift pattern found. For all distances, illusion scores ranged between 20 and 35, which on our continuous scale up to 100 resembles illusion scores previously found from 1 on a discrete scale to 3, e.g. \cite{Lloyd.2007}. Given this, we can assume that we were able to induce the illusion in every condition. Illusion scores and the proprioceptive drift, additionally, were not correlated. This supports the current debate that body-ownership illusions and the drift are two different, but related, processes \cite{Abdulkarim.2016}. The proprioceptive drift is an unconscious process - in contrast to the illusion, which is consciously accessible to the subjects. Hence, it might be possible that the predictive coding formulation in its unconscious form can explain drifting patterns, while it is not as such sufficient to explain body-ownership illusions. 

In contrast to \cite{Erro.2018}, we did not find a conclusive visual drift in the human experiment. From participants' personal communication, we know that many used visual landmarks to estimate the position of the rubber hand, but of the real hand also. Differences in this strategy for localization would not only account for the high variance we observed in the drift data, but also for the small magnitude of the proprioceptive drift - as compared to the values reported in other studies (e.g. \cite{Zopf.2010}). Beyond that, the method we used for localization is probably prone to high variance due to small mouse movements. Although we tried to account for that by repeating the localization ten times, more trials might be needed as performed in \cite{Samad.2015}. Arguably, however, the lack of visual drift in our study does not contradict the predictive coding scheme. Actually, some of our participants communicated that they experienced that the representation of both hands merged together. This is supported by the positive mean (14.16) of the actual control statement S10 ``It felt as if the rubber hand and my own right hand lay closer to each other", which was even higher than the mean response (-4.95) to the actual illusion statement S3 ``I felt as if the rubber hand were my hand". Further investigations, accounting for the variance in localization, are needed to support this conjecture.

The computational model presented here also generates predictions about the temporal dynamics of the rubber hand illusion. The constant accumulation of information resulting in an also accumulated drift of the body estimation (see \ref{fig:resultsrh:a2}) is comparable to findings from \cite{Tsakiris.2005} (see \ref{fig:overview}), who also found an accumulation of the drift over time in humans. To provide a finer temporal comparison, the dynamics of the human illusion should be further investigated.

\section{Conclusion}
\label{sec:conclusions}

We implemented the rubber hand illusion experiment on a multisensory robot. The perception of the real hand's position drifted towards the rubber hand, following a similar pattern in humans and the robot. We suggest that this proprioceptive drift resulted from a merging body estimation between both hands. This supports an account of the proprioceptive drift underlying body-ownership illusions in terms of the predictive coding scheme. Future work will address the mechanisms behind awareness of body-ownership.

\appendices

\section{Free energy gradient}
\label{sec:appendix:partialF}

\begin{equation}
p(\hat{x}|s_p,s_v) = p(s_p|\hat{x})p(s_v|\hat{x})p(s_{vt}|\hat{x}) p(\hat{x})
\end{equation}
Applying logarithms we get the negative free energy formulation:
\begin{equation}
F = \ln p(s_p|\hat{x}) + \ln p(s_v|\hat{x})  + \ln p(s_{vt}|\hat{x}) + \ln p(\hat{x})
\end{equation}
Substituting the probability distributions by their functions $f(.;.)$, and under the Laplace approximation \cite{friston2008hierarchical} and assuming normally distributed noise, we can compute the negative free energy as: 
\small
\begin{align}
\label{eq:freeenergyexample}
F =&   \ln f(s_p; g_p(x), \sigma_p) + \ln f(s_p; g_p(x), \sigma_p) + \ln f(x; \mu_x, \sigma_x) \nonumber\\
  =&  - \frac{(\hat{x} - \mu_x)^2}{2\sigma_x} + \nonumber\\
  & - \frac{(s_p - g_p(\hat{x}))^2}{2\sigma_p}   -\frac{(s_v - g_v(\hat{x}))^2}{2\sigma_v}  -\frac{(s_{vt} - g_{vt}(\hat{x}))^2}{2\sigma_{vt}}   
   \nonumber\\
  & \quad + \frac{1}{2} \left[  -\ln \sigma_x  - \ln \sigma_{s_p} - \ln \sigma_{s_v}  - \ln \sigma_{s_{vt}}\right] + C. 
\end{align}
\normalsize
In order to find $\hat{x}$ in a gradient-descent scheme we minimize Eq. \ref{eq:freeenergyexample} through the following differential equation:
\small
\begin{align}
\dot{x} =& - \frac{\hat{x}-\mu_x}{\sigma_x} +  \nonumber\\
&+ \frac{s_p - g_p(\hat{x})}{\sigma_p} g_p'(\hat{x}) +\frac{s_v - g_v(\hat{x})}{\sigma_v} g_v'(\hat{x}) +\frac{s_{vt} - g_{vt}(\hat{x})}{\sigma_{vt}} g'_{vt}(\hat{x})
\end{align}
\normalsize
In the case that $\hat{x}$ is equivalent to $g_{p}(x)$ like using the joint angles values directly as the body configuration, then  the proprioceptive error can be rewritten as: $s_p - \hat{x}$ and the gradient becomes 1.
Generalizing for $i$ sensors we finally have:
\begin{align}
\frac{\partial F}{\partial \hat{x}} = -\frac{(\hat{x} -\mu_x)}{\sigma_x} + \sum_i\frac{\partial{g_i(\hat{x})^T}}{\partial{\hat{x}}} \frac{s_i-g_i(\hat{x})}{\sigma_i}
\end{align}

\section{GP regression}
\label{sec:appendix:gp}
Given sensor samples $\overline{s}$ from the robot in several body configurations $\overline{x}$ and the covariance function $k(x_i,x_j)$, the \textit{training} is performed by computing the covariance matrix $K(X,X)$ on the collected data with noise $\sigma_n^2$:
\begin{equation}
\label{eq:covariance}
k_{ij} = \sigma_n^2\mathbf{I} + k(x_i,x_j) \quad | \forall i,j \in \overline{x}
\end{equation}

The \textit{prediction} of the sensory outcome $s$ given $x$ is then computed as \cite{rasmussen2005GPM}:
\begin{equation}
\label{eq:GPmean}
 g(\hat{x}) = k(\hat{x},X) K(X,X)^{-1} \overline{s}= k(\hat{x},X) \boldsymbol{\alpha}
\end{equation}
where $\boldsymbol{\alpha} = \text{choleski}(K)^T \backslash ( \text{choleski}(K) \backslash \overline{s})$.

Finally, in order to compute the gradient of the posterior $g(x)'$ we differentiate the kernel \cite{mchutchon2013differentiating}, and obtain its prediction analogously as Eq. \ref{eq:GPmean}:
\begin{align}
 g(\hat{x})' &= \frac{\partial k(\hat{x},X)}{\partial \hat{x}} K(X,X)^{-1} \overline{s} = \frac{\partial k(\hat{x},X)}{\partial \hat{x}} \boldsymbol{\alpha}
\end{align}
Using the squared exponential kernel with the Mahalanobis distance covariance function, the derivative becomes:
\begin{align}
\label{eq:gderivative}
g(\hat{x})' = -\Lambda^{-1} (\hat{x} - X)^T (k(\hat{x},X)^T \cdot \boldsymbol{\alpha})
\end{align}
where $\Lambda$ is a matrix where the diagonal is populated with the length scale for each dimension ($\text{diag}(1/l^2)$) and $\cdot$ is element-wise multiplication.

\addcontentsline{toc}{section}{References}
\bibliographystyle{IEEEtran}
\bibliography{pl,selfception,ninaRHI}

\end{document}